\documentclass[prb,twocolumn,showpacs,superscriptaddress,amsmath,amssymb]{revtex4}

\usepackage{graphicx}
\usepackage{latexsym}
\usepackage{amsmath}
\usepackage{amssymb}
\usepackage{amsfonts}
\usepackage{color}

\newcommand{\ds}{\displaystyle}

\begin{document}
\bibliographystyle{apsrev}

\title{Standard, inverse and triplet spin-valve effects in ${\bf F_1/S/F_2}$ systems}

\author{S.V. Mironov}
\affiliation{University
Bordeaux, LOMA UMR-CNRS 5798, F-33405 Talence Cedex, France}
\affiliation{Institute for Physics of Microstructures, Russian
Academy of Sciences, 603950 Nizhny Novgorod, GSP-105, Russia}
\author{A. Buzdin}
\affiliation{University
Bordeaux, LOMA UMR-CNRS 5798, F-33405 Talence Cedex, France}
\affiliation{Institut Universitaire de France, Paris, France}

\date{\today}
\begin{abstract}
We demonstrate that contrary to the common belief the critical temperature $T_c$ of clean ${\rm F_1/S/F_2}$ spin valves can depend non-monotonically on the angle between the magnetic moments of the ferromagnetic ${\rm F_1}$ and ${\rm F_2}$ layers. Depending on the system parameters the minimum of $T_c$ may correspond to parallel, antiparallel or non-collinear mutual orientation of magnetic moments. Such anomalous behavior can reveal itself only provided the ferromagnetic layers differ from each other and it completely disappears in the dirty limit.
\end{abstract}


\maketitle

\section{Introduction}

Multilayered ${\rm F_1/S/F_2}$ and ${\rm S/F_1/F_2}$ spin valves consisting of a superconductor (S) and two ferromagnetic layers (${\rm F_1}$ and ${\rm F_2}$) attract growing interest since they allow to manipulate the electronic transport in the superconductor by changing the magnetic state of the ferromagnets. The critical temperature $T_c$ of such systems depends on the angle $\theta$ between the magnetic moments of the ferromagnetic layers. Thus setting the temperature between minimum and maximum of $T_c$ and varying $\theta$ one can switch the system from normal to superconducting state.\cite{Beasley,Tagirov,Buzdin_FSF}

The physics behind the dependence $T_c\left(\theta\right)$ is closely related to the unusual proximity effect in superconductor/ferromagnet systems\cite{Buzdin_RMP,Golubov_RMP,Bergeret_RMP} which is known to be responsible also for Josephson $\pi-$junctions formation\cite{Buzdin_Pi,Ryazanov,Oboznov}, the oscillatory behavior of the critical temperature and the effective penetration depth of the S/F structures as functions of the F layer thickness\cite{Jiang,Zdravkov,Lemberger,Houzet}, the anomalous behavior of the electronic local density of states\cite{Buzdin_LDOS,Kontos,Baladie_DOS,Bergeret_DOS,Braude} as well as for the in-plane Fulde-Ferrell-Larkin-Ovchinnikov instability of the uniform superconducting state in multilayered S/F hybrids.\cite{Mironov1}

Even in the case of the s-wave superconductor the superconducting correlations produced in the ferromagnet contain spin-triplet components with the spin $S=1$ and different spin projection $S_z$. The decaying length of these components strongly depends on their spin structure. If the system contains only one ferromagnetic layer with uniform exchange field or several layers with collinear orientations of magnetic moments then the correlation function contains only triplet component with $S_z=0$, which is short-range and decays at a distance $\xi_f$ from the superconductor ($\xi_f$ is the superconducting correlation length in the ferromagnet which is typically much smaller than the correlation length $\xi_n$ in normal metal). At the same time if the magnetic moment in the F layers are non-collinear the long-range triplet components with $S_z=\pm 1$ arises, which are not affected by the exchange field and decay at a distance $\xi_n$. Another peculiarity of the proximity effect in S/F structures is the spatial oscillations of the correlation function inside the F layers, which originate from the Zeeman splitting of the Fermi surface in the ferromagnet.

The qualitative picture described above allows to explain most features of the dependencies $T_c\left(\theta\right)$ in superconducting spin valves. The key problem which was intensively studied during the past decade is the question about the mutual orientation of the magnetic moments (parallel or anti-parallel) corresponding to the higher critical temperature. One can naively expect that the critical temperature $T_c^{AP}$ for the anti-parallel orientation should always exceed the critical temperature $T_c^P$ for the parallel one ({\it standard spin-valve effect}) because for $\theta=\pi$ the average exchange field in the system is lower. At the same time detailed theoretical analysis of the dependencies $T_c\left(\theta\right)$ for ${\rm S/F_1/F_2}$ structures in the dirty limit show that it is not always the case.\cite{Fominov_SFF} The oscillatory behavior of the Cooper pair wave function inside the ferromagnets leads to the interference effects which depend strongly on the F layers thicknesses. As a result at certain ranges of thicknesses the value $\Delta T_c=T_c^{AP}-T_c^P$ becomes negative ({\it inverse spin-valve effect}).
Moreover it was shown that for $\theta\not= 0,\pi$ the appearance of long-range spin-triplet correlations opens an additional channel for the ``leakage" of Cooper pairs from the superconductor, which results in strong damping of the critical temperature. As a consequence in some region of system parameters the minimum of the dependence $T_c\left(\theta\right)$ corresponds to non-collinear orientation of magnetic moments ({\it triplet spin-valve effect}). The self-consistent numerical calculations performed on the basis of the Bogoliubov-de Gennes equations for clean ${\rm S/F_1/F_2}$ systems also show the possibility of the triplet spin-valve effect,\cite{Wu_SFF} but at the same time in these calculations only $T_c^{AP}>T_c^P$ behavior was observed.

The sign change of the value $\Delta T_c$ was observed experimentally for ${\rm CoO_x/Fe1/Cu/Fe2/In}$ and ${\rm V/Fe/V/Fe/CoO}$ multilayered system\cite{Leksin_1,Leksin_2,Leksin_3} while in the experiments with ${\rm V/Fe/V/Fe/CoO}$ structures\cite{Nowak_2} and the great variety of ${\rm S/F/N/F}$ systems with additional normal layer (N) between ferromagnets\cite{Nowak_1} only standard spin-valve effect was observed. At the same time experimental analysis of the full dependence $T_c\left(\theta\right)$ confirm the existence of the predicted triplet spin-valve effect.\cite{Leksin_2,Zdravkov_TSVE}

This pleasant agreement between theory and experiment breaks down for spin valves of the ${\rm F_1/S/F_2}$ type. The calculations performed for symmetric ${\rm F_1/S/F_2}$ structures (with identical ferromagnetic layers) predict only standard spin-valve effect in both dirty\cite{Tagirov,Buzdin_FSF,Baladie,You} and clean\cite{Bozovic,Halterman_PAP,Linder} limits [the results of Ref.~\onlinecite{Khusainov_Inverse}, where the inverse spin-valve effect was predicted, are questionable due to the possible inconsistency of the model (see Ref.~\onlinecite{Fominov_Kupriyanov})]. It was also shown that for F/S/F structures of atomic scale the anti-parallel orientation is usually favor the superconducting nucleation\cite{Daumens,Tollis} except the case of electron energy band inversion which may occur in strong ferromagnets due to the splitting of high energy band.\cite{Montiel} In Ref.~\onlinecite{Fominov_Kupriyanov} it is shown that for dirty symmetric spin valves with transparent S/F interfaces the dependencies $T_c\left(\theta\right)$ are always monotonically increasing for $0<\theta<\pi$. The behavior of the critical temperature in asymmetric dirty ${\rm F_1/S/F_2}$ spin valves was studied numerically in Refs.~\onlinecite{Lofwander} and \onlinecite{Cadden} where the authors found only monotonically increasing dependencies $T_c\left(\theta\right)$ with $T_c^{AP}>T_c^P$ even in the asymmetric case.

At the same time the experimental picture seems to be more rich. The majority of experiments for ${\rm F_1/S/F_2}$ structures show only the standard spin-valve effect\cite{Kinsey,Gu,Potenza,Wasterholt,Moraru_PRL,Moraru_PRB,Kim,Nowak_1,Luo} and monotonically increasing angular dependencies of the critical temperature,\cite{Zhu} which is in agreement with the theory. However in a number of experiments the inverse spin-valve effect was observed.\cite{Rusanov,Aarts,Steiner,Singh_APL,Singh_PRB,Leksin_Inverse} In some cases tuning the systems parameters leads to switching from standard to inverse spin-valve effect in resistance measurements.\cite{Zhu_Both,Hwang_Both}

All suggested explanations of the inverse spin-valve effect in ${\rm F_1/S/F_2}$ trilayers lay beyond the described proximity effect theory. The authors of Ref.~\onlinecite{Rusanov} pointed out that in case of strong ferromagnets the anti-parallel configuration of magnetic moments should lead to the accumulation of spin-polarized quasiparticles in the S layer due to relatively small probability of quasiparticle transmission from one F layer to another. Other authors suggested that the anomalous spin-valve effect may be caused by the stray field from domain walls in the ferromagnets\cite{Steiner,Flokstra} or by the dissipative vortex flow induced by the domain walls.\cite{Zhu_Both}

Thus it is agreed that the standard proximity effect in both clean and dirty ${\rm F_1/S/F_2}$ structures can cause only standard spin-valve effect which reveals in monotonically increasing dependencies of the critical temeprature of the angle between magnetic moments in ferromagnetic layers while the inverse spin-valve effect should be attributed to some other mechanisms.

In this paper we disprove this statement and show that clean ${\rm F_1/S/F_2}$ systems can reveal standard, inverse or triplet spin-valve effect due to the proximity effect only. The interference of quasiclassical trajectories, which is responsible to the spin-valve effect, is very sensitive to the width of ferromagnets. In particular if the F layers are identical only standard switching is possible while varying the width of one ferromagnet it is possible to tune the value $T_c^{AP}-T_c^P$ making it either positive or negative. At the same time in the dirty limit where such fine interference effects are absent the ${\rm F_1/S/F_2}$ structures are shown to reveal the standard spin-valve effect only.

The paper is organized as follows. In Sec.~\ref{Sec_FSF} we introduce our formalism and calculate the dependencies of the critical temperature on the mutual orientation of the magnetic moments of ${\rm F_1/S/F_2}$ spin valves in clean and dirty limits. In Sec.~\ref{Sec_Conc} we summarize our results.

\section{Spin-valve effect in F1/S/F2 structures}\label{Sec_FSF}

\begin{figure}[t!]
\includegraphics[width=0.25\textwidth]{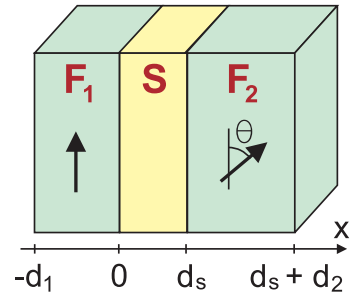}
\caption{(Color online) The sketch of multilayered ${\rm F_1/S/F_2}$ spin valve. The magnetic moments (shown with arrows) in the ferromagnetic layers form the angle $\theta$ with each other.} \label{Fig_SystemPicture}
\end{figure}

Let us consider the ${\rm F_1/S/F_2}$ spin valve, which is shown schematically In Fig.~\ref{Fig_SystemPicture}. The axis $x$ is chosen perpendicular to the layers interfaces. The exchange field ${\bf h}_1$ in the ${\rm F_1}$ layer is directed along the $z-$axis, while the exchange field ${\bf h}_2$ in the ${\rm F_2}$ layer lays in the $xz-$plane and form the angle $\theta$ with ${\bf h}_1$, so that $h_z=h_2 \cos\theta$ and $h_x=h_2 \sin\theta$.

In this case the quasiclassical anomalous Green function $\hat{f}$ can be represented in the form
\begin{equation}\label{f_def}
\hat{f}=f_s+\mathbf{f}_t\hat{\sigma},
\end{equation}
with only $f_s$, $f_{tz}$ and $f_{tx}$ components while $f_{ty}=0$.

We will assume the thickness of the S layer to be small enough, so that we can neglect the spatial variations of the superconducting gap function $\Delta$ across the superconductor. This assumption allows us to obtain the analytical expressions for the anomalous Green function both in clean (Sec.~\ref{Sec_FSF_Clean}) and dirty (\ref{Sec_FSF_Dirty}) limits.

\subsection{Clean limit}\label{Sec_FSF_Clean}

Let us assume that the exchange field in the ferromagnets is strong enough so that $h\tau\gg 1$ ($\tau$ is the elastic relaxation time). In this case to analyze the angular dependence of the critical temperature we use the system of linearized Eilenberger equations, which has the form\cite{Eschrig_ArXiv}
\begin{equation}\label{Eilenberger}
\left\{\begin{array}{l}{\ds v_F\cos\alpha\frac{\partial f_s}{\partial x}+2\omega_nf_s+2i{\bf h}{\bf f}_t=2\Delta,}\\{}\\{\ds v_F\cos\alpha\frac{\partial {\bf f}_t}{\partial x}+2\omega_n{\bf f}_t+2i{\bf h}f_s=0,}\end{array}\right.
\end{equation}
where $v_F$ is the absolute value of the Fermi velocity and $\alpha$ is the angle between the Fermi velocity and the $x-$axis.

The critical temperature $T_c$ of the system is determined by the self-consistency equation
\begin{equation}\label{SelfCons}
\Delta\ln \frac{T_c}{T_{c0}}+2\pi T_c\sum\limits_{n=0}^{\infty}\left(\frac{\Delta}{\omega_n}-\left<f_s\right>\right)=0,
\end{equation}
where $f_s$ is the spin-singlet component of the Green function inside the superconductor, $\left<...\right>$ denotes averaging over the angle $\alpha$, $\omega_n=\pi T_c\left(2n+1\right)$ is the Matsubara frequency, $T_{c0}$ is the critical temperature of the isolated superconducting layer.

It is convenient to introduce new functions
\begin{equation}\label{F_def}
\hat{F}^{\pm}\left(\alpha\right)=\frac{1}{2}\left[\hat{f}\left(\alpha\right)\pm\hat{f}^{+}\left(\alpha\right)\right],
\end{equation}
where $\hat{f}^{+}\left(\alpha\right)=\hat{f}\left(\pi-\alpha\right)$. Further we will use only the function $\hat{F}^{+}$ in the self-consistency equation, so we can consider only the angles for which $\cos\alpha>0$.

At S/F interfaces the functions $\hat{F}^{\pm}$ should be continuous while at the outer boundaries of the ferromagnets the function $\hat{F}^{-}$ should vanish due to the specular reflection of the quasiclassical trajectories. Thus, linear Eq.~(\ref{Eilenberger}) with the described boundary conditions allows a complete analytical solution. The corresponding calculations are presented in Appendix A. To present the expression for $F_s^+$ it is convenient first to introduce the values
\begin{equation}\label{Clean_RIT}
\begin{array}{c}{\ds T_j=\frac{v_s\cos\alpha_s}{2\omega_nd_s}\tanh\left[\frac{2\omega_nd_j}{v_j\cos\alpha_j}\right],}\\{}\\{\ds R_j=\frac{v_s\cos\alpha_s}{2\omega_nd_s}{\rm Re}\left\{\tanh\left[\frac{2\left(\omega_n+ih_j\right)d_j}{v_j\cos\alpha_j}\right]\right\},}\\{}\\{\ds I_j=\frac{v_s\cos\alpha_s}{2\omega_nd_s}{\rm Im}\left\{\tanh\left[\frac{2\left(\omega_n+ih_j\right)d_j}{v_j\cos\alpha_j}\right]\right\},}\end{array}
\end{equation}
where $j=1,2$ is the number of the ferromagnetic layer, $v_s$ and $v_j$ are the Fermi velocities in the superconductor and $j-$th ferromagnet respectively, $\alpha_s$ and $\alpha_j$ are the angles parameterizing the trajectories in S and F layers. Note that the values $T_j$ and $R_j$ are always positive while the values $I_j\propto\tan\left(2h_jd_j/v_j\cos\alpha_j\right)$ and their signs depend on $\alpha_j$.

The expression for $F_s^+$ can be represented in the following compact form:
\begin{equation}\label{FSF_F_Result}
F_s^+(\theta)=\frac{\Delta}{\omega_n}\frac{1}{\ds 1+R_1+R_2+Q(\theta)/W(\theta)},
\end{equation}
where
\begin{equation}\label{Clean_Q}
\begin{array}{c}{\ds Q(\theta)=\left(1+T_1+T_2\right)\left(I_1^2+2I_1I_2\cos\theta+I_2^2\right)+}\\{}\\{\ds + \left[I_1^2\left(R_2-T_2\right)+I_2^2\left(R_1-T_1\right)\right]\sin^2\theta}\end{array}
\end{equation}
and
\begin{equation}\label{Clean_W}
\begin{array}{c}{\ds W(\theta)=\left(1+T_1+T_2\right)\left(1+R_1+R_2\right)+}\\{}\\{\ds +\left(R_1-T_1\right)\left(R_2-T_2\right)\sin^2\theta.}\end{array}
\end{equation}

Let us analyze first the limiting cases of parallel and anti-parallel orientations of magnetic moments. Since $W(0)=W(\pi)$ the only difference between these cases is connected with the function $Q(\theta)$ which takes the form
\begin{equation}\label{Clean_Q_P_AP}
Q=\left(1+T_1+T_2\right)\left(I_1\pm I_2\right)^2,
\end{equation}
where the signs $``+"$ and $``-"$ correspond to parallel anti-parallel orientations respectively. From Eq.~(\ref{Clean_Q_P_AP}) one can easily see that if the ferromagnetic layers are identical then $Q(0)\geqslant Q(\pi)$ for all trajectories, and as a result $T_c^{AP}>T_c^P$. At the same time if the ferromagnetic layers differ from each other the interference effects become very important. On trajectories for which $I_1I_2>0$ the pair-breaking is stronger for parallel magnetic moments while on trajectories for which $I_1I_2<0$ the situation is opposite. The resulting effect is defined by the ratio between contributions from different trajectories.

To simplify the further procedure of averaging over the angle $\alpha_s$ in Eq.~(\ref{SelfCons}) we make several additional assumptions. Let us denote the Fermi momenta in superconductor and ferromagnets as $p_s$ and $p_f$ (we assume that $p_f$ is the same for both ferromagnets as well as $v_1=v_2\equiv v_f$). Then for a given quasiclassical trajectory the angles $\alpha_s$ and $\alpha_f$ inside the superconductor and ferromagnets satisfy the refraction law
\begin{equation}\label{Refraction}
p_s\sin\alpha_s=p_f\sin\alpha_f.
\end{equation}
Let us assume that $p_f\ll p_s$. In this case the pair-breaking is present only for trajectories with $\sin\alpha_s\leqslant p_f/p_s\ll 1$ since only these trajectories penetrate into the ferromagnet. Note that in principle the mismatch in the Fermi momenta can result in nonzero reflection factor at the S/F interfaces for trajectories with $\sin\alpha_s\leqslant p_f/p_s$. However we suppose that this factor changes very rapidly from zero to unity with the increase in $\alpha_s$ and the boundary conditions which were used to obtain Eq.~(\ref{FSF_F_Result}) are still valid. Then taking into account that for all trajectories which penetrate into the F layer $\cos\alpha_s\approx 1$, $\alpha_s\approx\left(p_f/p_s\right)\sin\alpha_f$, $d\alpha_s\approx\left(p_f/p_s\right)\cos\alpha_f d\alpha_f$ we obtain
\begin{equation}\label{Aver}
\left<...\right>=\int\limits_{0}^{\pi/2}...\sin\alpha_s d\alpha_s=\left(\frac{p_f}{p_s}\right)^2\int\limits_{1}^{\infty}...\frac{1}{\beta^3}d\beta,
\end{equation}
where $\beta=1/\cos\alpha_f$.

Since for $p_f\ll p_s$ only small part of the trajectories takes part into the pair-breaking process the critical temperature $T_c$ slightly differs from $T_{c0}$. In this case one can put $T_c\approx T_{c0}$ in the expression for $F_s^+$ and finally obtain that
\begin{equation}\label{Clean_Tc}
\frac{T_c}{T_{c0}}=1+\left(\frac{p_f}{p_s}\right)^2\sum \limits_{n=0}^{\infty}\int\limits_{1}^{\infty}\left(\frac{F_s^+}{2\pi T_{c0}\Delta}-\frac{1}{n+\frac{1}{2}}\right)\frac{d\beta}{\beta^3}.
\end{equation}

\begin{figure}[bt!]
\includegraphics[width=0.47\textwidth]{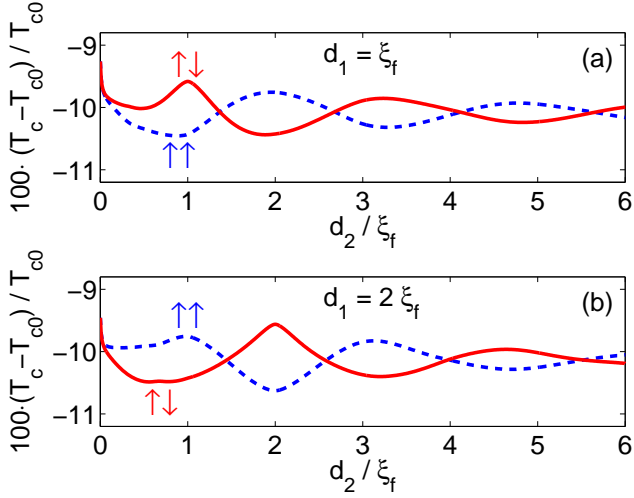}
\caption{(Color online) The dependencies of the critical temperature of the clean ${\rm F_1/S/F_2}$ spin valve for parallel (blue dashed curves marked $\uparrow\uparrow$) and anti-parallel (red solid curves marked $\uparrow\downarrow$) orientations of magnetic moments on the width $d_2$ of the ${\rm F_2}$ ferromagnetic layer. The width of the ${\rm F_1}$ layer takes the values (a) $d_1=\xi_f$ and (b) $d_1=2\xi_f$ (here $\xi_f=v_f/4\pi T_{c0}$). Other parameters are $h_1=h_2=2\pi T_{c0}$, $4\pi T_{c0}d_s/v_s=0.1$ and $p_f=0.2p_s$.} \label{Fig_Clean_PAP}
\end{figure}

\begin{figure}[bt!]
\includegraphics[width=0.47\textwidth]{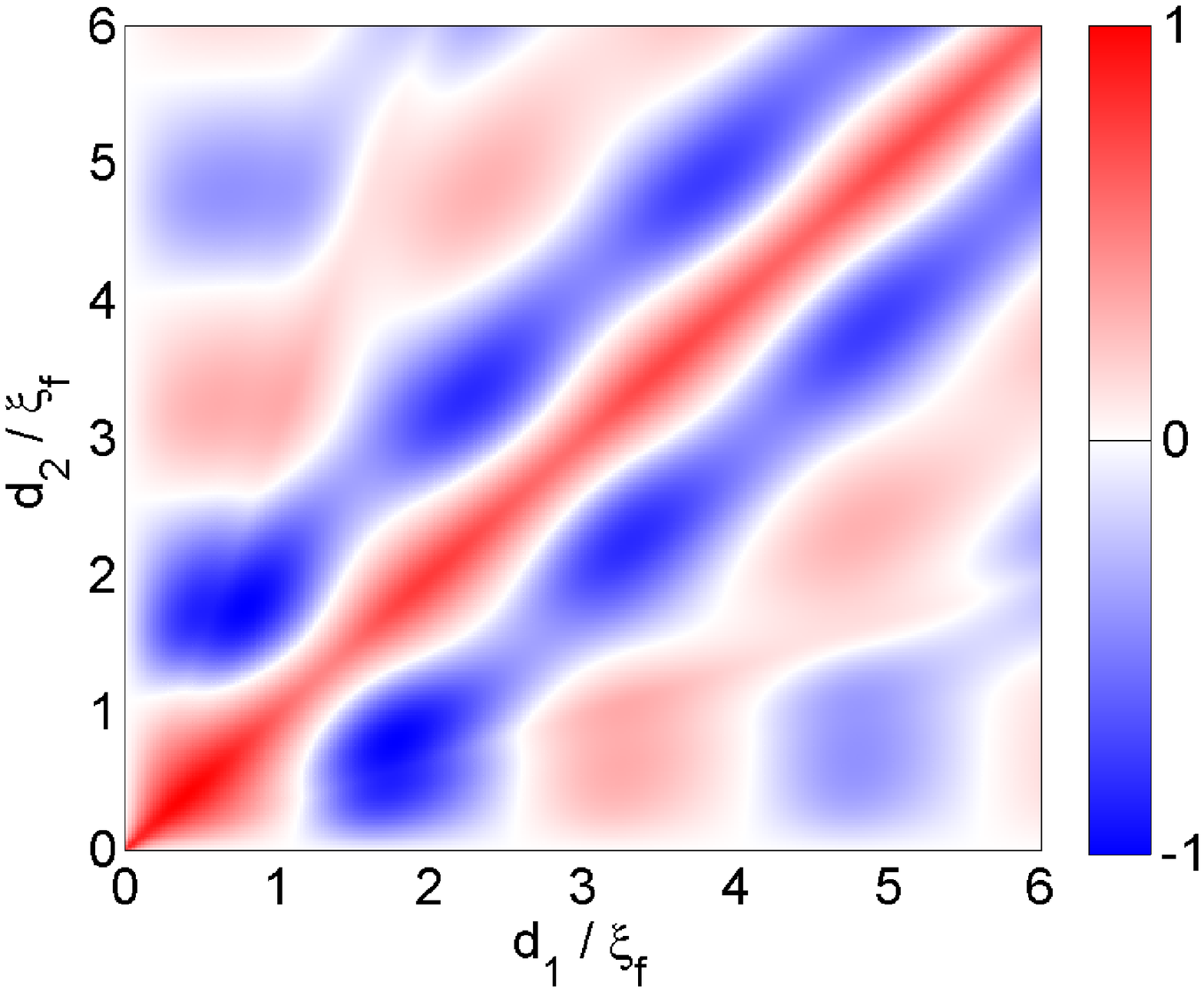}
\caption{(Color online) The dependence of the value $100\cdot\left(T_c^{AP}-T_c^P\right)/T_{c0}$ on the thicknesses of the ferromagnetic layers. The red (blue) areas correspond to the standard (inverse) spin-valve effect. The notations and parameters are the same as in Fig.~\ref{Fig_Clean_PAP}.} \label{Fig_SFS_Clean_Map}
\end{figure}

Using Eq.~(\ref{FSF_F_Result}) and performing numerical integration and summation over Matsubara frequencies in Eq.~(\ref{Clean_Tc}) we analyzed possible dependencies $T_c(\theta)$. The dependencies of $T_c^P$ ($\theta=0$) and $T_c^{AP}$ ($\theta=\pi$) on the thickness $d_2$ of the ${\rm F_2}$ ferromagnetic layer for two specific values of $d_1$ are shown in Fig.~\ref{Fig_Clean_PAP} (the estimated accuracy of numerical results is about $1\%$). One can see that depending on the thicknesses of the F layers either parallel or anti-parallel orientation of magnetic moments corresponds to the higher critical temperature. The full dependence of the value $\Delta T_c=T_c^{AP}-T_c^P$ on $d_1$ and $d_2$ is shown in Fig.~\ref{Fig_SFS_Clean_Map}, where red (blue) areas correspond to standard (inverse) spin-valve effect.

\begin{figure}[bt!]
\includegraphics[width=0.47\textwidth]{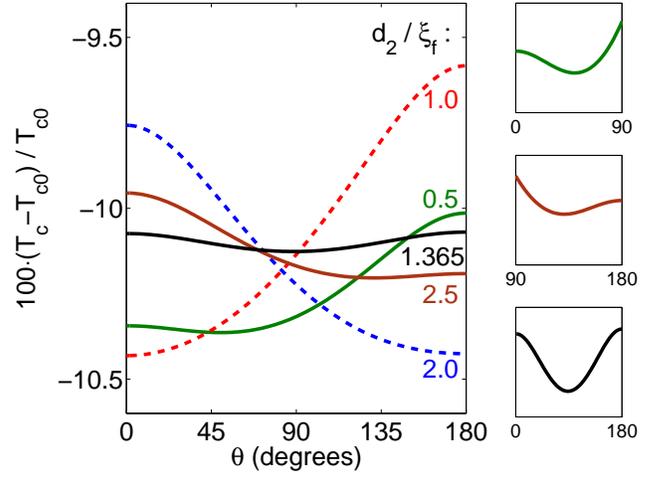}
\caption{(Color online) The dependencies of the critical temperature of the clean ${\rm F_1/S/F_2}$ spin valve on the angle $\theta$ between the magnetic moments in the ferromagnetic layers. Dashed red and blue curves are monotonic while solid green, brown and black curves demonstrate the triplet spin-valve effect (the enlarged fragments of these curves are shown in the insets). The parameters are $d_1=\xi_f$, $h_1=h_2=2\pi T_{c0}$, $4\pi T_{c0}d_s/v_s=0.1$, $p_f=0.2p_s$ and the values of the thickness $d_2$ of the $F_2$ layer are shown near the corresponding curves.} \label{Fig_Clean_Tc_Phi}
\end{figure}

Moreover the analysis of the angular dependencies of the critical temperature (see Fig.~\ref{Fig_Clean_Tc_Phi}) shows the possibility of the triplet spin-valve effect: for certain thicknesses of the F layers the minimum of $T_c$ corresponds to non-collinear orientation of magnetic moments. Thus depending on the parameters of ${\rm F_1/S/F_2}$ spin valve one can obtain standard, reverse or triplet switching.

Note also that the spin-valve effect in clean ${\rm F_1/S/F_2}$ structures reveals only if both ferrmagnetic layers have the finite width. If at least one of the layers (for example, ${\rm F_1}$) is infinite ($d_1\to\infty$) then in the above formulas $I_1\to 0$, $(R_1-T_1)\to 0$ and the critical temperature becomes independent on the angle $\theta$, which confirms the conclusion of Ref.~\onlinecite{Bozovic}.

\subsection{Dirty limit}\label{Sec_FSF_Dirty}

To calculate the critical temperature of the ${\rm F_1/S/F2}$ spin valve we use the linearized Usadel equation\cite{Eschrig_ArXiv}
\begin{equation}\label{Usadel}
\left\{\begin{array}{l}{\ds \frac{D}{2}\frac{\partial^2f_s}{\partial x^2}-\omega_nf_s- i{\bf h}{\bf f}_t=-\Delta,}\\{}\\{\ds \frac{D}{2}\frac{\partial^2{\bf f}_t}{\partial x^2}-\omega_n{\bf f}_t- i{\bf h}f_s=0,}\end{array}\right.
\end{equation}
where $D$ is the diffusion constant which can be different for different layers. The self-consistency equation has the form
\begin{equation}\label{SelfConsDirty}
\Delta\ln \frac{T_c}{T_{c0}}+2\pi T_c\sum\limits_{n=0}^{\infty}\left(\frac{\Delta}{\omega_n}-f_s\right)=0.
\end{equation}
At the outer boundaries of the ferromagnets one should demand $\partial\hat{f}/\partial x=0$ while at the S/F interfaces the boundary conditions read as\cite{Kupriyanov}
\begin{equation}\label{FSF_Dirty_BoundCond}
\frac{\partial \hat{f}^{(S)}}{\partial x}=\frac{\sigma_f}{\sigma_s}\frac{\partial \hat{f}^{(F)}}{\partial x}, ~~~\hat{f}^{(S)}=\hat{f}^{(F)}\mp \gamma_B\xi_s\frac{\partial \hat{f}^{(F)}}{\partial x}.
\end{equation}
Here $\sigma_s$ ($\sigma_f$) is the Drude conductivity of the superconductor (ferromagnet), $\xi_s=\sqrt{D_s/2\pi T_c}$ and the parameter $\gamma_B$ characterizes the interface transparency and is determined by the boundary resistance per unit area $R_b$ as $\gamma_B=R_b\sigma_f/\xi_s$. In the second condition in Eq.~(\ref{FSF_Dirty_BoundCond}) the sign $``-"$ ($``+"$) corresponds to the case when the normal from the superconductor to the ferromagnet is codirectional (contradirectional) with the $x$-axis.

The equations (\ref{Usadel}) together with the boundary conditions (\ref{FSF_Dirty_BoundCond}) allow us to obtain the analytical expression for spin-singlet Green function $f_s$ in the superconductor and calculate the critical temperature of the system. The details of the solution are presented in Appendix B. The resulting expression for $f_s$ formally coincides with Eq.~(\ref{FSF_F_Result}), where the functions $Q(\theta)$ and $W(\theta)$ are also defined by Eqs.~(\ref{Clean_Q}) and (\ref{Clean_W}), but the values $T_j$, $R_j$ and $I_j$ are redefined in the following way:
\begin{equation}\label{Dirty_RIT}
\begin{array}{l}{\ds T_j= \frac{\sigma_f}{\sigma_s}\frac{1}{q_s^2d_s}\frac{p_j\tanh\left(p_jd_j\right)}{1+\gamma_B\xi_sp_j\tanh\left(p_jd_j\right)},}\\{}\\{\ds R_j=\frac{\sigma_f}{\sigma_s}\frac{1}{q_s^2d_s}{\rm Re}\left\{\frac{q_j\tanh\left(q_jd_j\right)}{1+\gamma_B\xi_sq_j\tanh\left(q_jd_j\right)}\right\},} \\{}\\{\ds I_j=\frac{\sigma_f}{\sigma_s}\frac{1}{q_s^2d_s}{\rm Im}\left\{\frac{q_j\tanh\left(q_jd_j\right)}{1+\gamma_B\xi_sq_j\tanh\left(q_jd_j\right)}\right\},}\end{array}
\end{equation}
where $q_s=\sqrt{2\omega_n/D_s}$, $q_j=\sqrt{2\left(\omega_n+ih_j\right)/D_j}$, $p_j=\sqrt{2\omega_n/D_j}$ and $D_j$ is the diffusion constant in the $j-$th ferromagnetic layer. Note that in case of identical ferromagnetic layers the boundary conditions for the function $f_s$ in the superconducting layer coincide with the effective boundary conditions obtained in Ref.~\onlinecite{Fominov_Kupriyanov}.

In spite of the fact that the expressions for the spin-singlet component in the superconductor for clean and dirty limits are formally similar, the behavior of the critical temperature in these two cases is substantially different. In particular in the dirty limit all values $T_j$, $R_j$ and $I_j$ are positive. So from Eq.~(\ref{Clean_Q_P_AP}) one can see that $Q(0)\geqslant Q(\pi)$ and as a result $T_c^{AP}\geqslant T_c^P$ for arbitrary parameters of the ferromagnetic layers and S/F interfaces.

\section{Conclusion}\label{Sec_Conc}

Thus we calculated the critical temperature $T_c$ of ${\rm F_1/S/F_2}$ spin valves in the clean limit ($h\tau\gg 1$) and found that depending on the system parameters the minimum of $T_c$ may correspond to parallel, anti-parallel or non-collinear mutual orientation of the magnetic moments if ferromagnetic layers. The difference between $T_c$ for anti-parallel and parallel orientations is shown to be oscillatory function of the thicknesses of the F layers.

The interference effects which are responsible for the triplet and inverse spin-valve effects in ${\rm F_1/S/F_2}$ structures are sensitive to the disorder. We analyzed the behavior of $T_c$ in the dirty limit and show that for arbitrary parameters of the F layers and arbitrary transparencies of S/F interfaces the anti-parallel orientation of the magnetic moments provides more favorable conditions for the superconducting nucleation than the parallel one.
This statement generalize the similar conclusions obtained in Refs.~\onlinecite{Tagirov,Buzdin_FSF,Fominov_Kupriyanov} for a number of particular cases and the results of numerical calculations made in Refs.~\onlinecite{Lofwander,Cadden}.

Note that in the case of moderate disorder our results are formally not applicable. However we hope that for $h\tau\gtrsim 1$ the peculiarities of the spin-valve effect should be qualitatively similar to the ones in the clean limit and thus not only standard but also inverse and triplet switching can be observed.

\section*{ACKNOWLEDGMENTS}

The authors thank A.S.~Mel'nikov and A.V.~Samokhvalov for useful discussions and reading the manuscript. This work was
supported by the European IRSES program SIMTECH, French
ANR ``MASH," NanoSC COST Action MP1201, the Russian Foundation for Basic Research and
the Russian Presidential foundation (Grant No. SP-6340.2013.5).

\appendix

\section{Solution of the Eilenberger equations in the clean limit}\label{App_A}

In the superconductor the exchange field is zero and the general solution of Eq.~(\ref{Eilenberger}) has the form
\begin{equation}\label{Solution_S}
\begin{array}{l}{\ds F_s^+=\frac{\Delta}{\omega_n}+C_s^{(1)}\cosh\left(q_{s}x\right)+C_s^{(2)}\sinh\left(q_{s}x\right),}\\{}\\ {\ds {\bf F}_t^+={\bf C}_t^{(1)}\cosh\left(q_{s}x\right)+{\bf C}_t^{(2)}\sinh\left(q_{s}x\right),}\\{}\\{\ds \hat{F}^{-}=-\frac{1}{q_{s}}\frac{\partial\hat{F}^{+}}{\partial x},}\end{array}
\end{equation}
where $q_{s}=2\omega_n/v_s\cos\alpha_s$, $C_s^{(l)}$ and ${\bf C}_t^{(l)}$ ($l=1,2$) are the sets of scalar and vector constants (totally 6 scalar constants), which should be defined from the boundary conditions.

Considering the Eilenberger equations in ferromagnets one should put $\Delta=0$. To obtain the general solution of these equations in the $j-$th F layer ($j=1,2$) with the exchange field ${\bf h}_j$ it is convenient to represent the triplet part of the Green function as ${\bf F}_t^\pm={\bf F}_\parallel^\pm+{\bf F}_\perp^\pm$, where ${\bf F}_\parallel^\pm\parallel{\bf h}_j$ and ${\bf F}_\perp^\pm\perp{\bf h}_j$. Then if the boundary between the considered F layer and vacuum lays in the plane $x=x_j$ than the solution of Eq.~(\ref{Eilenberger}) which satisfies the boundary condition at $x=x_j$ has the form
\begin{equation}\label{Solution_F_General}
\begin{array}{l}{\ds F_s^+=A_j^{(1)}\cosh\left[q_j(x-x_j)\right]+A_j^{(2)}\cosh\left[q_j^*(x-x_j)\right],}\\{}\\{\ds F_s^-=-A_j^{(1)}\sinh\left[q_j(x-x_j)\right]-A_j^{(2)}\sinh\left[q_j^*(x-x_j)\right],}\\{}\\ {\ds F_\parallel^+=A_j^{(1)}\cosh\left[q_j(x-x_j)\right]-A_j^{(2)}\cosh\left[q_j^*(x-x_j)\right],}\\{}\\{\ds F_\parallel^-=-A_j^{(1)}\sinh\left[q_j(x-x_j)\right]+A_j^{(2)}\sinh\left[q_j^*(x-x_j)\right],}\\{}\\{\ds F_\perp^+=A_j^{(3)}\cosh\left[p_{j}(x-x_j)\right],}\\{}\\{\ds F_\perp^-=-A_j^{(3)}\sinh\left[p_{j}(x-x_j)\right],}\end{array}
\end{equation}
where $q_j=2(\omega_n+ih_j)/v_j\cos\alpha_j$, $p_j=2\omega_n/v_j\cos\alpha_j$, $A_j^{(l)}$ ($l=1,2,3$) are 3 constants and asterisk means the complex conjugation. Note that the components ${\bf F}_\perp^\pm$ are long-range and their decaying lengths do not depend on the exchange field.

Using the boundary conditions at S/F interfaces we obtain the system of 12 linear equations which allows to determine all unknown constants. In spite of all further calculations contain only algebraic transformations we will present one of possible ways of obtaining compact Eq.~(\ref{FSF_F_Result}) since it does not seem to be straightforward.

First let us mention that the assumption of the uniform $\Delta$ in the superconductor requires that $q_sd_s\ll 1$. This condition has to be fulfilled for $\omega_n$ less than the Debay frequency. So one can put $\cosh(q_sd_s)\approx 1$ and $\sinh(q_sd_s)\approx q_sd_s$. Another simplification can be made by redefining some of the unknown constants. Let us introduce new constants
\begin{equation}\label{AppI_NewConstants}
\begin{array}{l}{\ds B_j^{(1)}=A_j^{(1)}\cosh\left(q_jd_j\right),}\\{}\\ {\ds B_j^{(2)}=A_j^{(2)}\cosh\left(q_j^*d_j\right),}\\{}\\{\ds B_j^{(3)}=A_j^{(3)}\cosh\left(p_jd_j\right)}\end{array}
\end{equation}
and parameters
\begin{equation}\label{AppI_NewParameters}
\begin{array}{l}{\ds \kappa_j=\tanh\left(q_jd_j\right),}\\{}\\ {\ds \mu_j= \tanh\left(p_jd_j\right),}\\{}\\{\ds \lambda= q_sd_s.}\end{array}
\end{equation}
Then the system of equation takes the following form:
\begin{equation}\label{FSF_Clean_System_1}
\Delta/\omega_n+C_s^{(1)}=B_1^{(1)}+B_1^{(2)},
\end{equation}
\begin{equation}\label{FSF_Clean_System_2}
C_s^{(2)} = \kappa_1 B_1^{(1)}+\kappa_1^*B_1^{(2)},
\end{equation}
\begin{equation}\label{FSF_Clean_System_3}
C_z^{(1)}=B_1^{(1)}-B_1^{(2)},
\end{equation}
\begin{equation}\label{FSF_Clean_System_4}
C_z^{(2)}= \kappa_1 B_1^{(1)}-\kappa_1^*B_1^{(2)},
\end{equation}
\begin{equation}\label{FSF_Clean_System_5}
C_x^{(1)}=B_1^{(3)},
\end{equation}
\begin{equation}\label{FSF_Clean_System_6}
C_x^{(2)}=\mu_1 B_1^{(3)},
\end{equation}
\begin{equation}\label{FSF_Clean_System_7}
\Delta/\omega_n+C_s^{(1)}+\lambda C_s^{(2)}=B_2^{(1)}+B_2^{(2)},
\end{equation}
\begin{equation}\label{FSF_Clean_System_8}
-\lambda C_s^{(1)}-C_s^{(2)}= \kappa_2 B_2^{(1)}+\kappa_2^*B_2^{(2)},
\end{equation}
\begin{equation}\label{FSF_Clean_System_9}
C_z^{(1)}+\lambda C_z^{(2)}=\left(B_2^{(1)}-B_2^{(2)}\right)\cos\theta-B_2^{(3)}\sin\theta,
\end{equation}
\begin{equation}\label{FSF_Clean_System_10}
\lambda C_z^{(1)}+C_z^{(2)}=\left(\kappa_2^*B_2^{(2)}-\kappa_2B_2^{(1)}\right)\cos\theta+\mu_2B_2^{(3)}\sin\theta,
\end{equation}
\begin{equation}\label{FSF_Clean_System_11}
C_x^{(1)}+\lambda C_x^{(2)}=\left(B_2^{(1)}-B_2^{(2)}\right)\sin\theta+B_2^{(3)}\cos\theta,
\end{equation}
\begin{equation}\label{FSF_Clean_System_12}
\lambda C_x^{(1)}+C_x^{(2)}=\left(\kappa_2^*B_2^{(2)}-\kappa_2B_2^{(1)}\right)\sin\theta-\mu_2B_2^{(3)}\cos\theta.
\end{equation}
The equations (\ref{FSF_Clean_System_9})-(\ref{FSF_Clean_System_12}) can be transformed in a way that the right-hand side corresponding to the ${\rm F_2}$ layer would not depend on $\theta$. Physically this transformation means the rotation of the axes to make one of them directed along the exchange field in the ferromagnet. The result is
\begin{equation}\label{FSF_Clean_System_9a}
\left(C_x^{(1)}+\lambda C_x^{(2)}\right)\sin\theta+\left(C_z^{(1)}+\lambda C_z^{(2)} \right)\cos\theta=B_2^{(1)}-B_2^{(2)},
\end{equation}
\begin{equation}\label{FSF_Clean_System_10a}
\left(C_x^{(1)}+\lambda C_x^{(2)}\right)\cos\theta-\left(C_z^{(1)}+\lambda C_z^{(2)} \right)\sin\theta=B_2^{(3)},
\end{equation}
\begin{equation}\label{FSF_Clean_System_11a}
\begin{array}{l}{\ds
\left(\lambda C_x^{(1)}+C_x^{(2)}\right)\sin\theta+\left(\lambda C_z^{(1)}+C_z^{(2)} \right)\cos\theta=}\\{}\\{\ds ~~~~~~~~~~~~~~~~~~~~~~~~~~~~~~~~~~=-\kappa_2B_2^{(1)}+\kappa_2^*B_2^{(2)},}\end{array}
\end{equation}
\begin{equation}\label{FSF_Clean_System_12a}
\left(\lambda C_x^{(1)}+C_x^{(2)}\right)\cos\theta-\left(\lambda C_z^{(1)}+C_z^{(2)} \right)\cos\theta=-\mu_2B_2^{(3)}.
\end{equation}
Then getting rid of the constants corresponding to the superconductor, introducing new values $G=B_1^{(1)}+B_1^{(2)}=B_2^{(1)}+B_2^{(2)}$, $U=i\left(B_1^{(1)}-B_1^{(2)}\right)$, $V=i\left(B_2^{(1)}-B_2^{(2)}\right)$ and expressing all other constants by means of $G$, $U$ and $V$ we obtain the following system of 3 equations:
\begin{equation}\label{FSF_Clean_FinalEq_1}
\left[\lambda+{\rm Re}\left(\kappa_1+\kappa_2\right)\right]G+{\rm Im}\left(\kappa_1\right)U+{\rm Im}\left(\kappa_2\right)V=\lambda,
\end{equation}
\begin{equation}\label{FSF_Clean_FinalEq_2}
\begin{array}{l}{\ds
\left[{\rm Im}\left(\kappa_2\right)+{\rm Im}\left(\kappa_1\right)\cos\theta\right]G-\left[{\rm Re}\left(\kappa_1\right)-\mu_1\right]U\cos\theta-}\\{}\\{\ds ~~~~~~~~~~~~~~~~~~~~~~~~~~~~~~~~-\left[\lambda+\mu_1+{\rm Re}\left(\kappa_2\right)\right]V=0,}\end{array}
\end{equation}
\begin{equation}\label{FSF_Clean_FinalEq_3}
\begin{array}{l}{\ds
\left[{\rm Im}\left(\kappa_1\right)+{\rm Im}\left(\kappa_2\right)\cos\theta\right]G-\left[{\rm Re}\left(\kappa_2\right)-\mu_2\right]U\cos\theta-}\\{}\\{\ds ~~~~~~~~~~~~~~~~~~~~~~~~~~~~~~~~-\left[\lambda+\mu_2+{\rm Re}\left(\kappa_1\right)\right]V=0.}\end{array}
\end{equation}
Finally solving this system of equations (we need the expression for $G$ only) and making algebraic simplifications  we obtain Eq.~(\ref{FSF_F_Result}) with $T_j=\mu_j/\lambda$, $R_j={\rm Re}\left(\kappa_j\right)/\lambda$ and $I_j={\rm Im}\left(\kappa_j\right)/\lambda$.

\section{Solution of the Usadel equation in the dirty limit}\label{App_B}

In the superconductor the solution of Eq.~(\ref{Usadel}) has the following form:
\begin{equation}\label{Dirty_Solution_S}
\begin{array}{l}{\ds f_s=\frac{\Delta}{\omega_n}+C_s^{(1)}\cosh\left(q_{s}x\right)+C_s^{(2)}\sinh\left(q_{s}x\right),}\\{}\\ {\ds {\bf f}_t={\bf C}_t^{(1)}\cosh\left(q_{s}x\right)+{\bf C}_t^{(2)}\sinh\left(q_{s}x\right),}\end{array}
\end{equation}
where $q_s=\sqrt{2\omega_n/D_s}$, $D_s$ is the diffusion constant in the superconductor.

As for the case of the clean limit (see Appendix~\ref{App_A}) to write the solution inside the ferromagnet it is convenient to substend the triplet component of the Green function along and across the exchange field vector: ${\bf f}_t={\bf f}_\parallel+{\bf f}_\perp$. Then if the outer boundary of the ferromagnet corresponds to the plane $x=x_j$ then the solution of the Usadel equation is
\begin{equation}\label{Dirty_Solution_F}
\begin{array}{l}{\ds f_s=A_j^{(1)}\cosh\left[q_j(x-x_j)\right]+A_j^{(2)}\cosh\left[q_j^*(x-x_j)\right],}\\{}\\ {\ds f_\parallel=A_j^{(1)}\cosh\left[q_j(x-x_j)\right]-A_j^{(2)}\cosh\left[q_j^*(x-x_j)\right],}\\{}\\{\ds f_\perp=A_j^{(3)}\cosh\left[p_{j}(x-x_j)\right],}\end{array}
\end{equation}
where $q_j=\sqrt{2\left(\omega_n+ih_j\right)/D_j}$ and $q_j=\sqrt{2\omega_n/D_j}$ ($D_j$ is the diffusion constant in the $j-$th ferromagnet).

Substituting Eq.~(\ref{Dirty_Solution_S}) and (\ref{Dirty_Solution_F}) into the boundary conditions (\ref{FSF_Dirty_BoundCond}) we obtain the system of 12 linear equations. The procedure of the solution has much in common with the solution of the Eilenberger equations presented in Appendix~\ref{App_A}.

The assumption of the constant gap function $\Delta$ in the superconductor is valid provided $q_sd_s\ll1$. Then one may put $\cosh\left(q_sd_s\right)\approx 1$ and $\sinh\left(q_sd_s\right)\approx q_sd_s$.

Let us redefine the constants corresponding to the ferromagnetic layers and introduce new constants
\begin{equation}\label{AppB_NewConstants}
\begin{array}{l}{\ds B_j^{(1)}=A_j^{(1)}\left[\cosh\left(q_jd_j\right)+\gamma_B\xi_sq_j\sinh\left(q_jd_j\right)\right],}\\{}\\ {\ds B_j^{(2)}=A_j^{(2)}\left[\cosh\left(q_j^*d_j\right)+\gamma_B\xi_sq_j^*\sinh\left(q_j^*d_j\right)\right],}\\{}\\{\ds B_j^{(3)}=A_j^{(3)}\left[\cosh\left(p_jd_j\right)+\gamma_B\xi_sp_j\sinh\left(p_jd_j\right)\right]}\end{array}
\end{equation}
and parameters
\begin{equation}\label{AppB_NewParameters}
\begin{array}{l}{\ds \kappa_j=\frac{\sigma_f}{\sigma_s}\frac{q_j}{q_s}\frac{\tanh\left(q_jd_j\right)}{1+\gamma_B\xi_sq_j\tanh\left(q_jd_j\right)},}\\{}\\ {\ds \mu_j= \frac{\sigma_f}{\sigma_s}\frac{p_j}{q_s}\frac{\tanh\left(p_jd_j\right)}{1+\gamma_B\xi_sp_j\tanh\left(p_jd_j\right)},}\\{}\\{\ds \lambda= q_sd_s.}\end{array}
\end{equation}
In this case the system of equation becomes formally identical to the system (\ref{FSF_Clean_System_1})-(\ref{FSF_Clean_System_12}) and thus it has the same solution but with new parameters (\ref{AppB_NewParameters}).

\end{document}